\documentclass[conference]{IEEEtran}
\IEEEoverridecommandlockouts
\usepackage{cite}
\usepackage{amsmath,amssymb,amsfonts}
\usepackage{algorithmic}
\usepackage{amsmath}
\usepackage{mathtools}
\usepackage{booktabs}
\usepackage{mathrsfs}
\usepackage{makecell}
\usepackage{multirow}
\usepackage{graphicx}
\usepackage{textcomp}
\usepackage{xcolor}
\usepackage{algorithm}
\usepackage{upgreek}

\def\BibTeX{{\rm B\kern-.05em{\sc i\kern-.025em b}\kern-.08em
    T\kern-.1667em\lower.7ex\hbox{E}\kern-.125emX}}
\begin{document}

\title{A 0.21-ps FOM Capacitor-Less Analog LDO with 
Dual-Range Load Current for Biomedical  Applications\\

}

\author{\IEEEauthorblockN{Yasemin Engür and Mahsa Shoaran}
\IEEEauthorblockA{\text{Institute of Electrical and Micro Engineering, and Neuro-X Institute} \\
\text{École Polytechnique Fédérale de Lausanne (EPFL), Lausanne, Switzerland }\\
Email: yasemin.engur@epfl.ch; mahsa.shoaran@epfl.ch\vspace{-4mm}
}
}

\maketitle

\begin{abstract}

This paper presents an output capacitor-less low-dropout regulator (LDO) with a bias switching scheme for biomedical applications with dual-range load currents. Power optimization is crucial for systems with multiple activation modes such as neural interfaces, IoT and edge devices with varying load currents. 
To enable rapid switching between low and high current states, a flipped voltage follower (FVF) configuration is utilized, along with a super source follower buffer to drive the power transistor. Two feedback loops and an on-chip compensation capacitor ($\text{C}_\text{C}$) maintain the stability of the regulator under various load conditions. 
The LDO was implemented in a 65nm CMOS process with 1.5V input and 1.2V output voltages. The measured quiescent current is as low as 3$\upmu$A and 50$\upmu$A for the 0-500$\upmu$A and 5-15mA load current ranges, respectively. An undershoot voltage of 100mV is observed when the load current switches from 0 to 15mA within 80ns, with a maximum current efficiency of 99.98\%. Our design achieved a low Figure-of-Merit of 0.21ps, outperforming state-of-the-art analog LDOs.

\end{abstract}

\vspace{-2mm}
\section{Introduction}
Low-dropout regulators (LDOs) play a critical role in power management circuits across a wide range of applications. They can provide multiple voltage levels required for powering a variety of circuits, including biomedical systems, processors, cameras, mobile and IoT devices \cite{b1,b2,b3}. In such applications, the LDO must deliver a stable power source with a rapid transient response and minimal power consumption. While digital regulators offer a fast transient response and low power, their accuracy and power supply rejection (PSR) are limited by intrinsic quantization error \cite{b1}. In contrast, analog LDOs provide well-regulated noise-free output voltages with higher bandwidth PSR, making them suitable for noise-sensitive low-quiescent-current analog/bio/RF applications \cite{BioCAS-LDO,ChangJSSC}.

Prior research attempted to improve the performance of analog regulators, focusing on metrics such as quiescent current, load transient response, settling and response time, and PSR. For example, adaptive biasing was proposed in \cite{b3} to achieve a fast  transient response, at the cost of a high  quiescent current of 133\text{$\mu$}A. 
The capacitive feedforward ripple cancellation method in \cite{b2} can enhance supply noise rejection performance, but requires a 1\text{$\mu$}F off-chip capacitor, increasing the occupied area and vulnerability to bonding wire parasitic effects. Alternatively, output capacitor-less LDO regulators were presented in \cite{YanLu,b4,A-SSCC2022} to provide a broadband PSR, at the expense of a high quiescent current. 
Achieving an optimal regulator performance requires a balance between the design trade-offs. Ideally, the LDO should provide minimal settling time, low quiescent current and small voltage fluctuations, and occupy a compact area \cite{b5,b6}.

\begin{figure}[t]
\centering
\includegraphics[width=1\columnwidth]{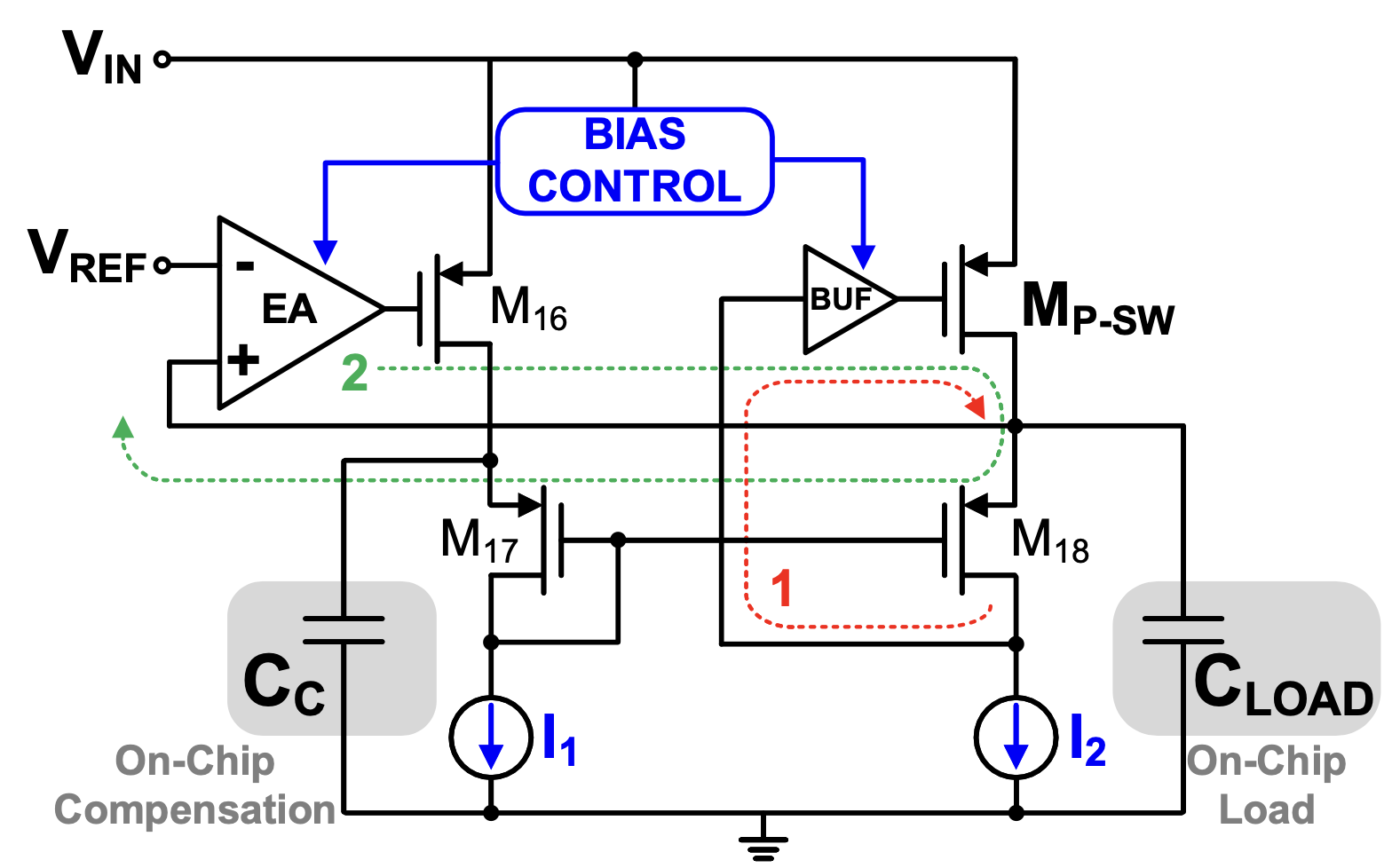}
\vspace{-8mm}  
\caption{Top-level architecture of the proposed dual-current-range capacitor-less  low-dropout regulator.} \vspace{-5mm}  
\label{fig1}
\end{figure}

\begin{figure*}[t]
\centering
\includegraphics[width=2\columnwidth]{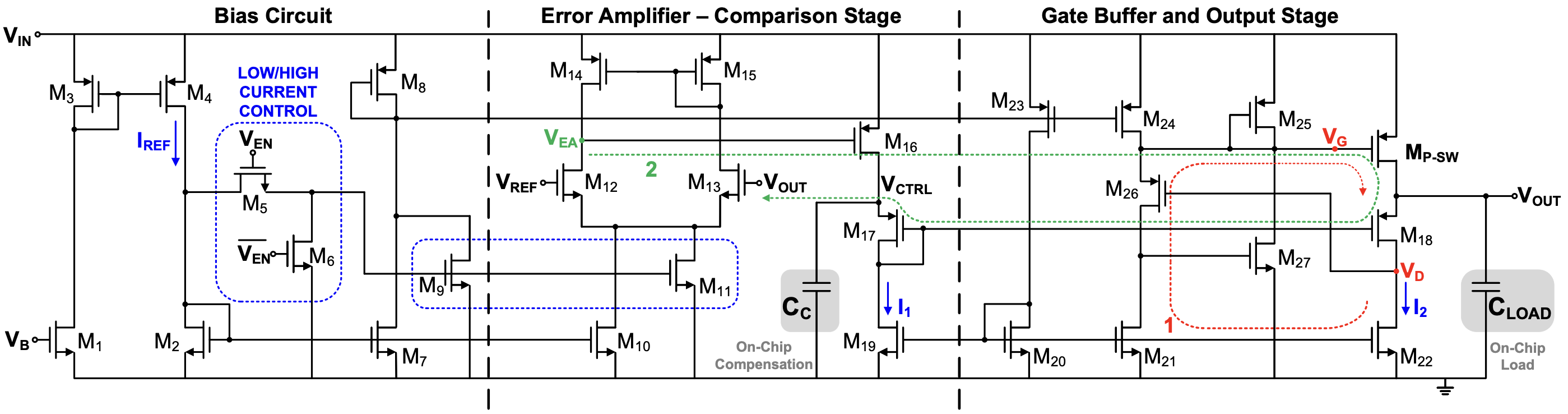}
\vspace{-3mm}  
\caption{The circuit diagram of the proposed dual-current output capacitor-less flipped voltage follower LDO.} \vspace{-3mm}  
\label{fig2}
\end{figure*}

Different system-on-chips (SoCs) may require specific activation modes tailored to their particular operation.  For instance, a closed-loop neural interface  continuously records and processes neural signals, while  stimulation is only triggered upon detection of  disease symptoms \cite{shoaran2016hardware}, resulting in a rapid and temporary increase in current and power consumption \cite{b7,b8,yoo2021neural}. Similarly, multi-channel bio-signal amplifiers can be activated  altogether or on-demand, with power consumption increasing with the number of active channels. In IoT sensors and edge  devices, the load current can significantly vary during sleep/wake-up cycles and upon event detection, making the LDO design challenging to support multiple operation modes~\cite{YanLu2,b9}.
To address this issue, we present a capacitor-less LDO with a new bias switching scheme. Our approach supports both high- and low-current regulation modes to provide a reliable supply voltage for integrated circuits operating under varying load currents (e.g., biomedical and IoT). 

The overall architecture of the proposed regulator is illustrated in Fig.~\ref{fig1}. The bias control block is responsible for high-low current mode switching. To achieve a swift transition from low to high current mode, we have adopted a flipped voltage follower (FVF) topology \cite{b4}. In addition, we have increased the load drive capability by incorporating a super source follower buffer, while ensuring circuit stability through the use of two feedback loops and an on-chip compensation capacitance.
An output-pole-dominant configuration has been implemented using an on-chip load capacitance to effectively suppress the output noise and glitches, while extending the lower end of the load current range.

This paper is organized as follows. Section II describes the implementation of the proposed capacitor-less LDO that supports dual-range load currents. Section III presents the experimental results, while Section IV concludes the paper.

\section{Circuit Implementation}
The schematic circuit diagram for the dual-current FVF regulator is presented in Fig.~\ref{fig2}. The circuit consists of three primary stages: 1) a low-high current mode-controlling bias circuit; 2) a two-stage error amplifier for the comparison stage; and 3) a PMOS output voltage switch with a gate buffer and a load capacitance $\text{C}_\text{LOAD}$.

\subsection{Bias Switching for Dual-Range Load Currents}

Current requirements of a circuit can vary widely (e.g., from \text{$\mu$}As to mAs) depending on the application and the specific circuit blocks that are activated. The proposed bias switching circuit generates a stable output voltage for two distinct current intervals, referred to low and high current modes.

\begin{figure}[t]
\centering
\includegraphics[width=1\columnwidth]{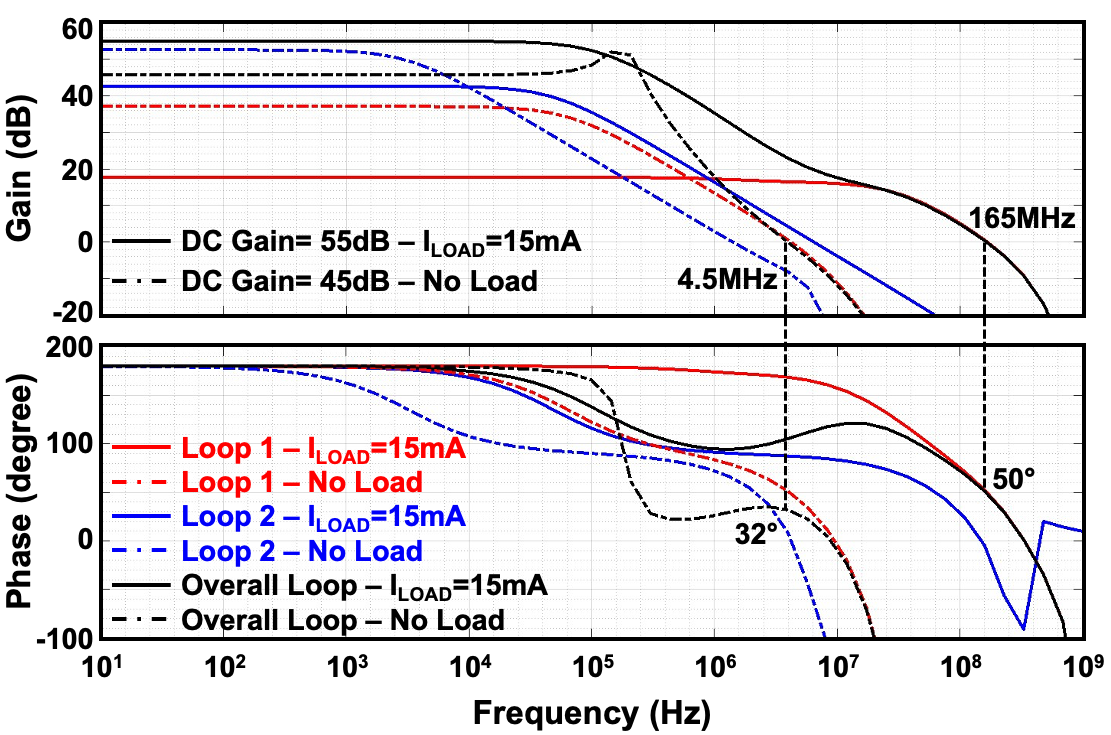}
\vspace{-5mm}  
\caption{The simulated frequency response of the feedback loops for No load and 15mA load current, with $\text{V}_{\text{IN}}$=1.5V and $\text{V}_{\text{OUT}}$=1.2V. The red/blue legends on the bottom plot  refer to the corresponding curves in both sub-figures.} 
\vspace{-5mm}  
\label{fig3}
\end{figure}

For the low load currents, the FVF regulator is biased by  voltage $\text{V}_\text{B}$  in low current mode,  generating the reference current  $\text{I}_\text{REF}$. The $\text{I}_\text{REF}$ passes through  $\text{M}_\text{2}$ and $\text{M}_\text{4}$ and is mirrored by  $\text{M}_\text{7}$/$\text{M}_\text{10}$ to bias the error amplifier. During low current mode, the load current remains within the 0-500\text{$\mu$}A range and the $\text{V}_\text{EN}$ signal remains low. 
To prevent current leakage, $\text{M}_\text{6}$ connects the gates of  $\text{M}_\text{9}$ and $\text{M}_\text{11}$ to the ground. Meanwhile,  $\text{M}_\text{23}$/$\text{M}_\text{24}$ mirror the gate voltage of $\text{M}_\text{8}$ to bias the rest of the regulator. 
With this configuration, the measured quiescent current of the regulator is only 3\text{$\mu$}A, while the LDO regulates the output voltage to 1.2V from a 1.5V supply ($\text{V}_\text{IN}$). 

As the load requires more current, the voltage at $\text{V}_\text{EN}$ rises, which causes the regulator to switch to the high current mode. $\text{M}_\text{5}$ creates a second biasing path by connecting to the gates of both $\text{M}_\text{9}$ and $\text{M}_{11}$, increasing the overall current of the regulator while maintaining a constant reference current $\text{I}_\text{REF}$.
The quiescent current in the  high current mode varies within 50-110\text{$\mu$}A to meet the load current requirements, which can range from 5 to 15mA. 
An output dominant pole configuration is attained by using a super source follower buffer to shift the pole of $\text{M}_{\text{P-SW}}$ to higher frequencies. 
While the buffer causes a slight increase in power consumption in the high current mode, the bias switching circuit switches to the low current mode when a high load current is not required, effectively mitigating this increase.

\subsection{Stability Analysis} 
Dual-range current adjustment with bias switching reduces the LDO's power consumption. To ensure the stability of the regulator in both low and high current modes, two feedback loops are utilized. Loop 2 provides the output voltage for one of the inputs of the error amplifier. 
The two-stage error amplifier (EA) compares the output voltage to $\text{V}_\text{REF}$ and generates  $\text{V}_\text{EA}$, thus completing Loop 2.  Following the voltage comparison in the first stage, the second stage of the error amplifier produces the $\text{V}_\text{CTRL}$ signal. This two-stage process is essential in generating a clear gate voltage for the $\text{M}_\text{17}$-$\text{M}_\text{18}$ pair.
The Loop 2 feedback mechanism can be dominated by the poles of the two-stage EA, which include both the internal and output poles at $\text{V}_\text{EA}$ and $\text{V}_\text{CTRL}$, respectively. To ensure good PSR in mid-range frequencies, the dominant pole is set at the output pole of the EA. This is achieved by adding a compensation capacitor $\text{C}_\text{C}$ to the node $\text{V}_\text{CTRL}$. The analysis of  Loop 2 is performed by breaking the feedback loop at the node $\text{V}_\text{EA}$.
\\
\indent In the output stage, the gate of the PMOS switch $\text{M}_{\text{P-SW}}$ is controlled by the  flipped voltage follower configuration.  Loop 1 employs a gate buffer that connects the drain of $\text{M}_{\text{22}}$ ($\text{V}_{\text{D}}$) to the gate of $\text{M}_{\text{P-SW}}$ ($\text{V}_{\text{G}}$), resulting in a reduction of input capacitance at $\text{V}_{\text{D}}$ and a lower output impedance at $\text{V}_{\text{G}}$. This, in turn, moves the gate pole of $\text{p}_{\text{gate}}$ towards higher frequencies, thus improving the stability  of the circuit.
Indeed, this approach ensures that system stability is always maintained, regardless of the output pole's dependence on load current, enabling reliable operation under different load conditions.
In order to improve the regulator's speed, the circuit employs a current ratio of $\text{I}_{\text{1}}$:$\text{I}_{\text{2}}$ equal to 1:4. Additionally, the diode-connected transistor $\text{M}_{\text{25}}$ is utilized to improve the buffer's pull-up capability and transient response, further enhancing the regulator's  performance. Loop 1 stability analysis is performed by isolating it from the rest of the circuit. Figure \ref{fig3} shows the simulations of the DC gain and phase characteristics for Loop 1, Loop 2, and the overall stability under the low and high current  operations (i.e., at  No load  and a load current of 15mA). Loop 1 exhibits high bandwidth with lower DC gain, making it essential for high-speed switching, while the slower loop (i.e., Loop 2) is crucial for achieving accurate output voltages. Overall, the regulator's stability is guaranteed with unity-gain bandwidths of 4.5MHz and 165MHz for the low and high current modes, respectively. 

\begin{figure}[t]
\centering
\includegraphics[width=0.7\columnwidth]{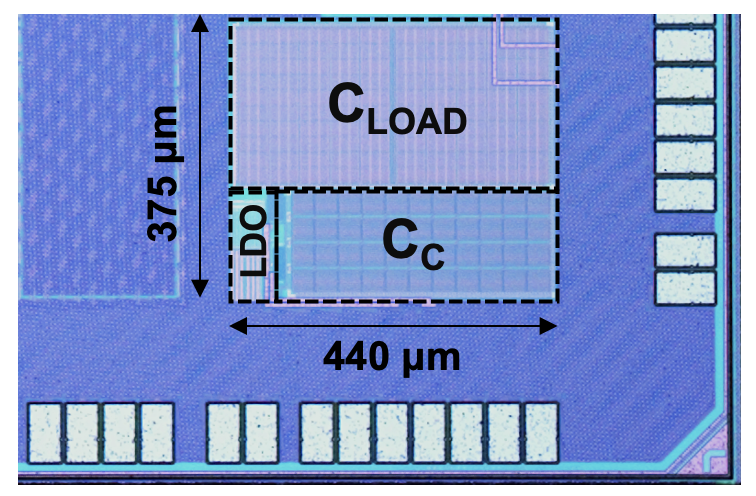}
\vspace{-3mm}  
\caption{Chip micrograph of the proposed LDO in 65nm TSMC LP process.} \vspace{-3mm}  
\label{fig4}
\end{figure}

\begin{figure}[t]
\centering
\includegraphics[width=1\columnwidth]{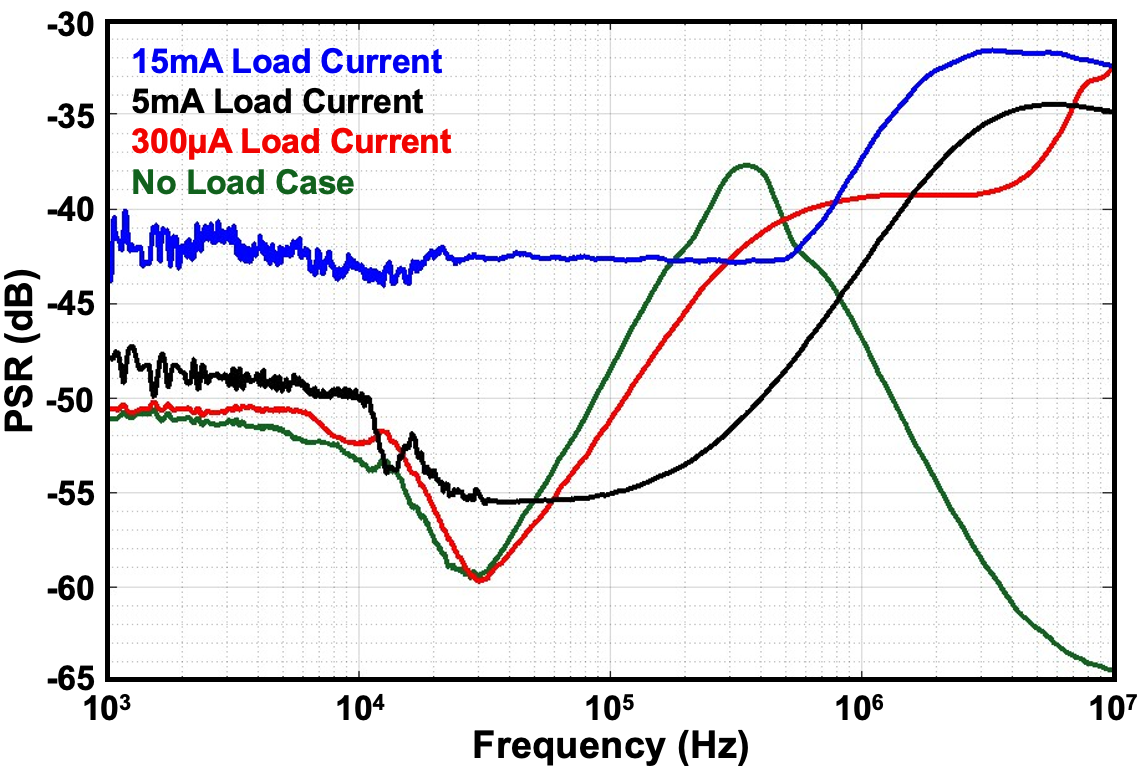}
\vspace{-5mm}  
\caption{The measured PSR of the proposed LDO at different load currents.} \vspace{-3mm}  
\label{fig5}
\end{figure}

\begin{figure}[t]
\centering
\includegraphics[width=1\columnwidth]{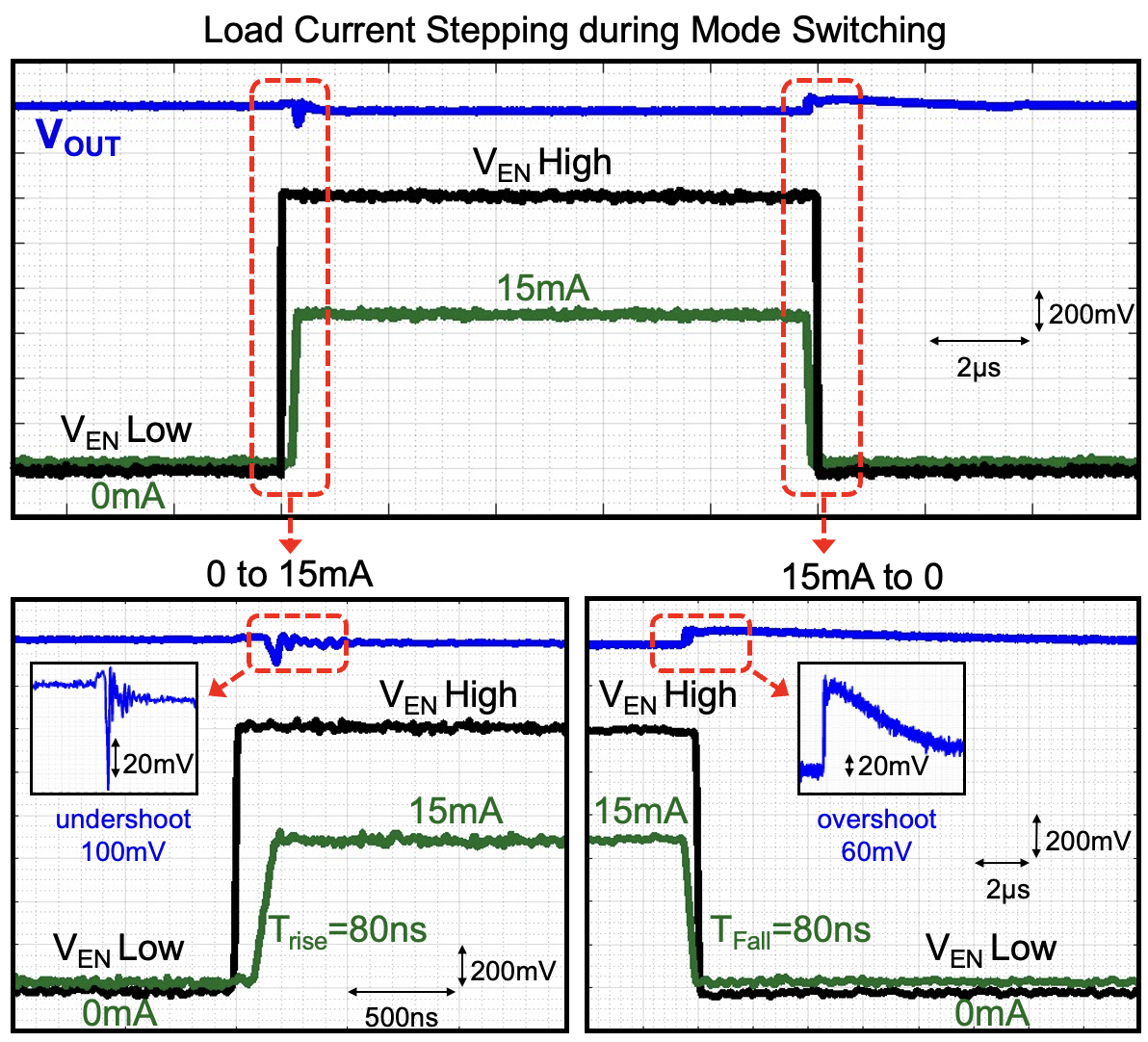}
\vspace{-5mm}  
\caption{The measured transient response for $\text{V}_{\text{IN}}$=1.5V, $\text{V}_{\text{OUT}}$=1.2V, and the load current stepping from 0 to 15mA during mode switching.} \vspace{-5mm}  
\label{fig6}
\end{figure}

\section{Measurement Results}
The proposed capacitor-less analog regulator was fabricated in a 65nm TSMC LP process, as depicted in Fig. \ref{fig4}. The active chip area is 0.165mm$^2$ including the on-chip compensation($\text{C}_{\text{C}}$ = 40pF) and the load capacitors ($\text{C}_{\text{LOAD}}$ = 160pF). To ensure the desired capacitance matching, we opted to employ on-chip metal-insulator-metal (MIM)  capacitors rather than stacking multiple capacitor types (such as MOS, MOM, and MIM as in \cite{b4}). Although this design choice resulted in a slight increase in chip area, it was deemed necessary to ensure adequate capacitance matching for achieving high PSR, which is crucial for reliable LDO operation. This dual-current LDO consumes 3\text{$\mu$}A of quiescent current in the low current mode. When employing externally controlled mode switching, the measured quiescent current varies between 50\text{$\mu$}A and 110\text{$\mu$}A, depending on the load current  that ranges from 5mA to 15mA.

\begin{table*}[t]
  \centering  
  \vspace{-5mm}    
  \caption{Comparison with the State-of-the-Art Analog Regulators.}\vspace{-3mm} 
  \includegraphics[width=1.7\columnwidth]{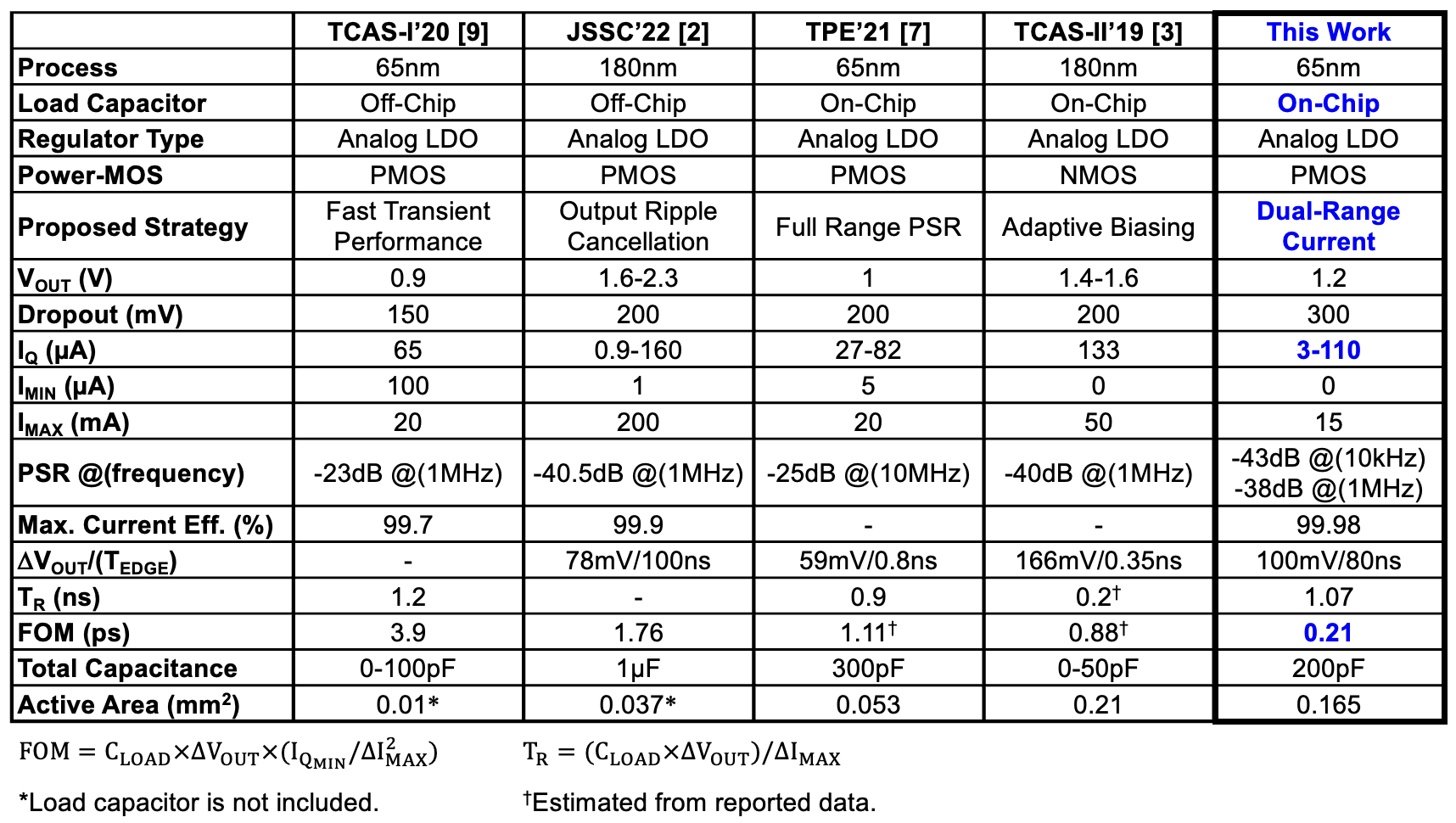}\vspace{-3mm}  
  \label{CompTable}
\end{table*}

\begin{figure}[t]
\centering
\includegraphics[width=1\columnwidth]{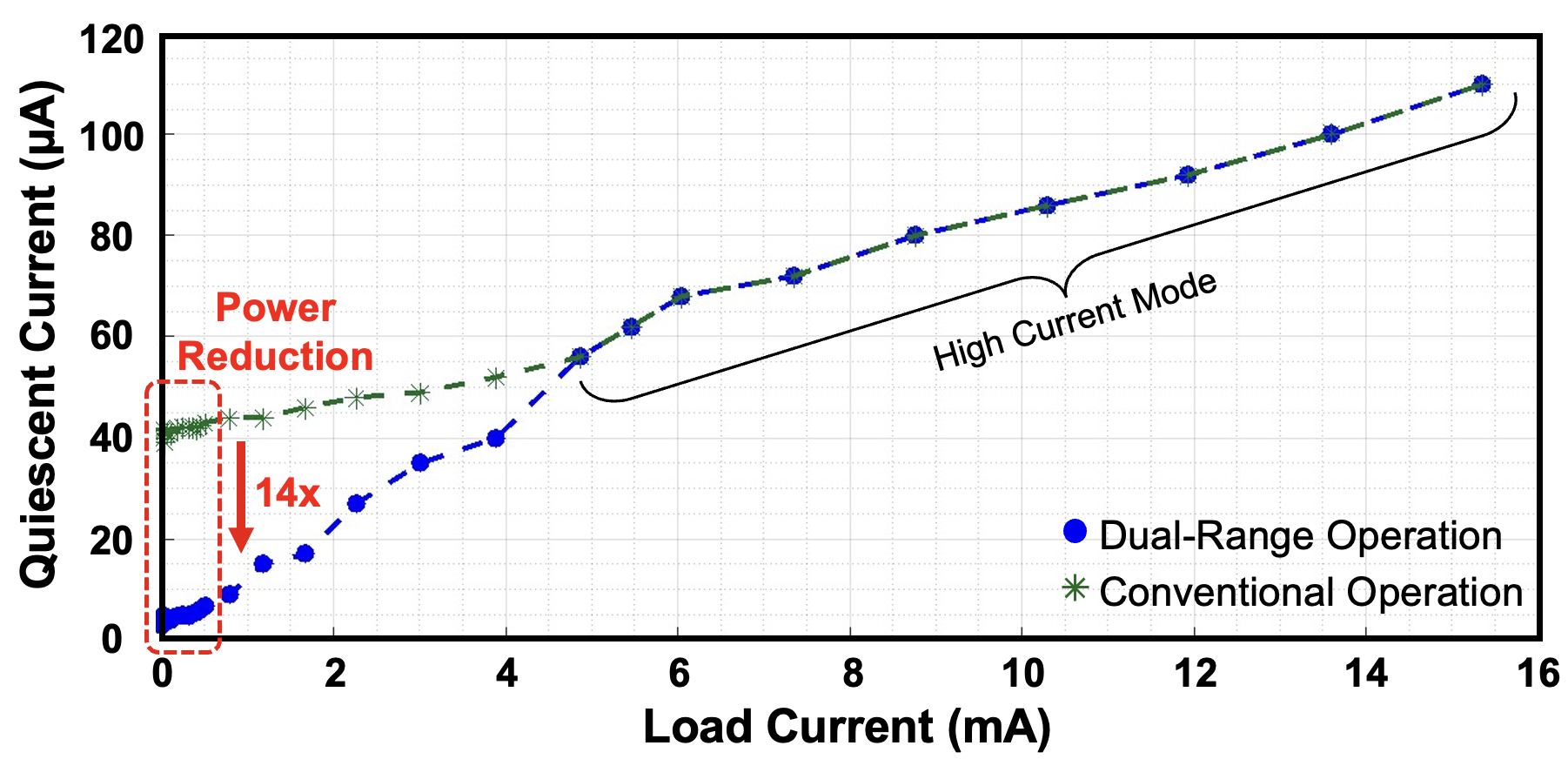}
\vspace{-5mm}  
\caption{The measured quiescent currents for the dual-range  versus the conventional LDO architectures.} \vspace{-5mm}  
\label{fig7}
\end{figure}

In the proposed regulator, the error amplifier dominates the PSR behaviour at low frequencies, while the on-chip load capacitor suppresses the higher frequency noise. Figure~\ref{fig5} shows the measured PSR in 1kHz to 10MHz frequency range under varying load currents.  
The results demonstrate that the LDO can effectively attenuate supply noise by 40dB up to 10kHz, which is the typical frequency range for most biomedical and sensor interface applications \cite{shin2022neuraltree}. The regulator achieves over 32dB of suppression up to 10MHz in both low and high current modes, highlighting its robustness in mitigating supply noise across a broad  range of frequencies.

Figure~\ref{fig6} shows the measured transient response for $\text{V}_{\text{IN}}$=1.5V, $\text{V}_{\text{OUT}}$=1.2V, and the on-chip load current changing from 0 to 15mA. As soon as the $\text{V}_{\text{EN}}$ signal changes from low to high, mode switching begins and lasts until the high current demand continues. During this transition, undershoot and overshoot voltages of 100mV and 60mV were observed, with the rising and falling edge times of 80ns, respectively. 
The output voltage settles in less than 500ns during the transition from low to high current mode. Furthermore, the regulator's power consumption is considerably reduced. If the circuit had continuously operated in the high current mode, it would have required a quiescent current of 42\text{$\mu$}A to maintain functionality, as illustrated in Fig. \ref{fig7}. With the proposed dual-range current approach, the regulator can function with a quiescent current of only 3\text{$\mu$}A, thus  decreasing the quiescent current by 14$\times$ compared to the conventional approach and improving the system efficiency. Particularly, circuits that predominantly operate under low-current modes can achieve substantial power savings by utilizing this approach.

The performance  of the proposed low-dropout regulator is compared with the state-of-the-art analog LDOs in Table \ref{CompTable}. 
The proposed regulator achieves the lowest quiescent current using the dual-range current approach, with the exception of \cite{b2}. In \cite{b2}, the large off-chip output capacitor eliminates the need for pole shifting to stabilize the LDO, at the cost of excessively large area. The flipped voltage follower configuration along with the pull-up diode led to fast switching, with a low measured response time of 1.07ns. 
Moreover, the proposed regulator presents a PSR performance that is on par with or comparable to   prior works, while achieving the lowest FOM of 0.21ps, outperforming the state-of-the-art analog LDOs.

\vspace{-1mm}

\section{Conclusion}

In this paper, a capacitor-less analog regulator with a new mode switching approach was presented. The dual-range load current operation effectively reduces the LDO's power consumption. To enable high-speed switching, a flipped voltage follower configuration along with a super source follower buffer were used. 
The design was fabricated in 65nm CMOS process, occupying an active area of 0.165mm$^2$. The proposed LDO operates with  quiescent currents of 3\text{$\mu$}A and 50-110\text{$\mu$}A in low and high current modes, respectively. 
This dual-range regulator achieved an outstanding FOM of 0.21ps, with a transient response time of 1.07ns. The measured PSR is less than 40dB up to 10kHz. 
The achieved specifications demonstrate that this design is well-suited for applications that demand fast settling, low noise, and varying load currents, such as biomedical and sensor interface applications.

\vspace{-2mm}

\bibliographystyle{IEEEtran.bst}
\bibliography{yaseminengur}

\begin{thebibliography}{10}
\providecommand{\url}[1]{#1}
\csname url@samestyle\endcsname
\providecommand{\newblock}{\relax}
\providecommand{\bibinfo}[2]{#2}
\providecommand{\BIBentrySTDinterwordspacing}{\spaceskip=0pt\relax}
\providecommand{\BIBentryALTinterwordstretchfactor}{4}
\providecommand{\BIBentryALTinterwordspacing}{\spaceskip=\fontdimen2\font plus
\BIBentryALTinterwordstretchfactor\fontdimen3\font minus \fontdimen4\font\relax}
\providecommand{\BIBforeignlanguage}[2]{{%
\expandafter\ifx\csname l@#1\endcsname\relax
\typeout{** WARNING: IEEEtran.bst: No hyphenation pattern has been}%
\typeout{** loaded for the language `#1'. Using the pattern for}%
\typeout{** the default language instead.}%
\else
\language=\csname l@#1\endcsname
\fi
#2}}
\providecommand{\BIBdecl}{\relax}
\BIBdecl

\bibitem{b1}
J.-G. Kang, J.~Park, M.-G. Jeong, and C.~Yoo, ``Digital low-dropout regulator with voltage-controlled oscillator based control,'' \emph{IEEE Transactions on Power Electronics}, vol.~37, no.~6, pp. 6951--6961, 2022.

\bibitem{b2}
T.~Guo, W.~Kang, and J.~Roh, ``A 0.9-$\upmu$$\text{A}$ quiescent current high $\text{PSRR}$ low dropout regulator using a capacitive feed-forward ripple cancellation technique,'' \emph{IEEE Journal of Solid-State Circuits}, vol.~57, no.~10, pp. 3139--3149, 2022.

\bibitem{b3}
D.~Mandal, C.~Desai, B.~Bakkaloglu, and S.~Kiaei, ``Adaptively biased output cap-less $\text{NMOS}$ $\text{LDO}$ with 19ns settling time,'' \emph{IEEE Transactions on Circuits and Systems II: Express Briefs}, vol.~66, no.~2, pp. 167--171, 2019.

\bibitem{BioCAS-LDO}
Y.~Huang, F.~Kong, J.~Freeman, and L.~Najafizadeh, ``A low drop-out regulator for subcutaneous electrical stimulation of nanofibers used in muscle prosthesis,'' in \emph{2015 IEEE Biomedical Circuits and Systems Conference (BioCAS)}, 2015, pp. 1--4.

\bibitem{ChangJSSC}
J.~Jiang, W.~Shu, and J.~S. Chang, ``A 65-nm $\text{CMOS}$ low dropout regulator featuring $>$60-$\text{dB}$ $\text{PSRR}$ over 10-$\text{MHz}$ frequency range and 100-m$\text{A}$ load current range,'' \emph{IEEE Journal of Solid-State Circuits}, vol.~53, no.~8, pp. 2331--2342, 2018.

\bibitem{YanLu}
Y.~Lu, Y.~Wang, Q.~Pan, W.-H. Ki, and C.~P. Yue, ``A fully-integrated low-dropout regulator with full-spectrum power supply rejection,'' \emph{IEEE Transactions on Circuits and Systems I: Regular Papers}, vol.~62, no.~3, pp. 707--716, 2015.

\bibitem{b4}
G.~Cai, Y.~Lu, C.~Zhan, and R.~P. Martins, ``A fully integrated $\text{FVF}$ $\text{LDO}$ with enhanced full-spectrum power supply rejection,'' \emph{IEEE Transactions on Power Electronics}, vol.~36, no.~4, pp. 4326--4337, 2021.

\bibitem{A-SSCC2022}
H.~Park, W.~Jung, M.~Kim, and H.-M. Lee, ``A fast-transient and wide-range output capacitor-less $\text{NMOS}$ $\text{LDO}$ regulator with adaptive-gain nested miller compensation and pre-emphasis inverse biasing,'' in \emph{2022 IEEE Asian Solid-State Circuits Conference (A-SSCC)}, 2022, pp. 6--8.

\bibitem{b5}
N.~Liu and D.~Chen, ``A transient-enhanced output-capacitorless $\text{LDO}$ with fast local loop and overshoot detection,'' \emph{IEEE Transactions on Circuits and Systems I: Regular Papers}, vol.~67, no.~10, pp. 3422--3432, 2020.

\bibitem{b6}
H.-J. Choi, J.-M. Cho, H.-J. Park, and S.-W. Hong, ``An output capacitor-less low-dropout regulator using a wide-range single-stage gain-boosted error amplifier and a frequency-dependent buffer with a total compensation capacitance of 2.5p$\text{F}$ in 0.5µm $\text{CMOS}$,'' in \emph{2021 IEEE Asian Solid-State Circuits Conference (A-SSCC)}, 2021, pp. 1--3.

\bibitem{shoaran2016hardware}
M.~Shoaran, M.~Farivar, and A.~Emami, ``Hardware-friendly seizure detection with a boosted ensemble of shallow decision trees,'' in \emph{2016 38th Annual International Conference of the IEEE Engineering in Medicine and Biology Society (EMBC)}.\hskip 1em plus 0.5em minus 0.4em\relax IEEE, 2016, pp. 1826--1829.

\bibitem{b7}
U.~Shin, L.~Somappa, C.~Ding, Y.~Vyza, B.~Zhu, A.~Trouillet, S.~P. Lacour, and M.~Shoaran, ``A 256-channel 0.227$\upmu$$\text{J}$/class versatile brain activity classification and closed-loop neuromodulation soc with 0.004mm$^2$-1.51µ$\text{W}$/channel fast-settling highly multiplexed mixed-signal front-end,'' in \emph{2022 IEEE International Solid- State Circuits Conference (ISSCC)}, vol.~65, 2022, pp. 338--340.

\bibitem{b8}
B.~Zhu, U.~Shin, and M.~Shoaran, ``Closed-loop neural prostheses with on-chip intelligence: A review and a low-latency machine learning model for brain state detection,'' \emph{IEEE Transactions on Biomedical Circuits and Systems}, vol.~15, no.~5, pp. 877--897, 2021.

\bibitem{yoo2021neural}
J.~Yoo and M.~Shoaran, ``Neural interface systems with on-device computing: Machine learning and neuromorphic architectures,'' \emph{Current Opinion in Biotechnology}, vol.~72, pp. 95--101, 2021.

\bibitem{YanLu2}
Y.~Lu, ``Digitally assisted low dropout regulator design for low duty cycle $\text{IoT}$ applications,'' in \emph{2016 IEEE Asia Pacific Conference on Circuits and Systems (APCCAS)}, 2016, pp. 33--36.

\bibitem{b9}
J.~Zhao, Y.~Gao, T.-T. Zhang, H.~Son, and C.-H. Heng, ``A 310-$\text{nA}$ quiescent current 3-fs-$\text{FoM}$ fully integrated capacitorless time-domain $\text{LDO}$ with event-driven charge pump and feedforward transient enhancement,'' \emph{IEEE Journal of Solid-State Circuits}, vol.~56, no.~10, pp. 2924--2933, 2021.

\bibitem{shin2022neuraltree}
U.~Shin, C.~Ding, B.~Zhu, Y.~Vyza, A.~Trouillet, E.~C. Revol, S.~P. Lacour, and M.~Shoaran, ``Neuraltree: A 256-channel 0.227-$\mu$j/class versatile neural activity classification and closed-loop neuromodulation soc,'' \emph{IEEE Journal of Solid-State Circuits}, vol.~57, no.~11, pp. 3243--3257, 2022.

\end{thebibliography}

\vspace{12pt}
\end{document}